\documentclass[showpacs,floatfix,twocolumn,prl,citeautoscript,superscriptaddress]{revtex4-1}
\usepackage{graphicx}
\usepackage{bm}
\usepackage{amsmath}
\usepackage{amssymb}

\newcommand{\qav}[1]{\left \langle {#1} \right \rangle}

\begin{document}
\title{Universal decay cascade model for dynamic quantum dot initialization}
\author{Vyacheslavs Kashcheyevs}
\affiliation{Faculty of Computing, University of Latvia, Riga LV-1586, Latvia}
\affiliation{Faculty of Physics
and Mathematics, University of Latvia, Riga LV-1002, Latvia}
\author{Bernd Kaestner}
\affiliation{Physikalisch-Technische Bundesanstalt, Bundesallee
100, 38116 Braunschweig, Germany}

\begin{abstract}
Dynamic quantum dots can be formed by time-dependent
electrostatic potentials in nanoelectronic devices, such as
gate- or surface-acoustic-wave-driven electron pumps. Ability
to control the number of captured electrons with high precision
is required  for applications in fundamental metrology and
quantum information processing.
In this work we propose and quantify a scheme to initialize quantum
dots with a controllable number of electrons.
It is based on the stochastic decrease in the
electron number of a shrinking dynamic quantum dot
and is described by a nuclear decay cascade model
with ``isotopes'' being different charge states of the dot.
Unlike the natural nuclei, the artificial confinement
is time-dependent and tunable, so the probability
distribution for the final ``stable isotopes''
depends on the external gate voltage.
We derive an explicit fitting formula to extract the sequence
of decay rate ratios from the measurements of averaged
current in a periodically driven device.
This provides a device-specific fingerprint
which allows to compare different devices and
architectures, and predict the upper limits of initialization
accuracy from low precision measurements.
\end{abstract}

\maketitle

Single electron charging effects have attracted much interest
since the proposal of single electronics~\cite{Averin1991} and
the possibility to fabricate nanoscale structures. In
particular, quantum dots (QD) connected to leads have been a
standard model system for many years to study single charges in
so called artificial atoms. Dynamic QDs, which are repeatedly
formed and manipulated by time-varying confining potentials
appear in particular
in structures proposed to study quantum
information~\cite{loss1QIBI, elzerman2004, barnes1,
feve2007, kataoka2007, hayashi2008}.
One of the issues to be addressed is the decoupling
of the QD from the environment and at the same time
allowing the fast initialization with a controllable number of electrons.
A suitable method has been demonstrated by
Kataoka~\emph{et al.}~\cite{kataoka2007} which
uses pulses of surface acoustic waves (SAWs) to populate or depopulate
a QD that is well isolated from electron reservoirs.
While this mechanism is not yet fully understood,
a more conventional approach~\cite{LiuNiu93}
is adiabatic decoupling of the QD from electron reservoir,
keeping the voltage on the QD fixed in a Coulomb blockade valley
separating the discrete charge states.
This strategy is limited by
(i) non-adiabatic excitation of the QD~\cite{Flensberg1999}
due to necessary finite decoupling rate,
and (ii)  experimental difficulties in tuning the lead-dot coupling to zero
without disturbing the electrostatic potential on the QD, $\varphi$.

Here we propose and quantify an alternative scheme to achieve
initialization by allowing non-equilibrium relaxation
(backtunneling) from a QD being raised energetically
above the Fermi level during the decoupling process.
This process is known to play a role in several types of
non-adiabatic current generation devices~\cite{Aizin1998,
Robinson2001,blumenthal2007a,miyamoto2008,Leicht2009}.
We identify scale separation in integrated time-dependent
electron escape rates between the subsequent charge states
as a precondition for low-dispersion initialization.
In our decay-cascade model a voltage parameter $V$
shifts the hierarchy of the decay rates
and thereby tunes the target number of electrons, $n_0$,
and the corresponding error $\langle (n- n_0)^2 \rangle$.
The minimal error is then given by
the particular QD implementation and fixed,
for instance, by QD geometry.
We proceed to analyze the results of recent non-adiabatic current generation experiments \cite{fujiwara2008,kaestner2009a,Janssen2001}
and find them to be promising realizations of the proposed initialization scheme.
Based on the decay cascade model predictions we propose an indirect way to measure
the initialization accuracy in these devices by extracting the decay rate hierarchy fingerprint
from low precision measurements of their current-voltage characteristics. Finally, possible
strategies for accuracy improvement are discussed based on the ways
charging energy, temperature and barrier design affect the electron escape rates.

\begin{figure}
  \includegraphics[width=7cm,clip]{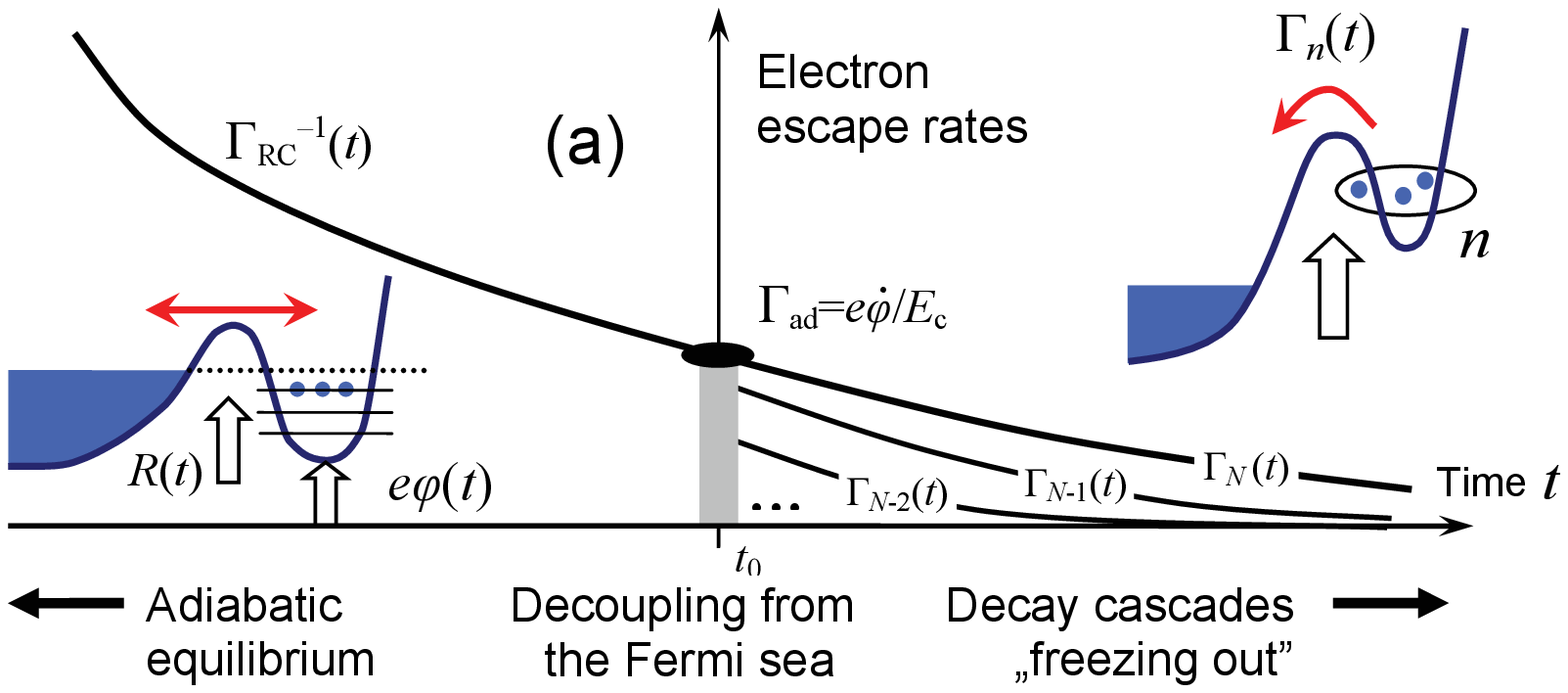} \\
  \includegraphics[width=7cm,clip]{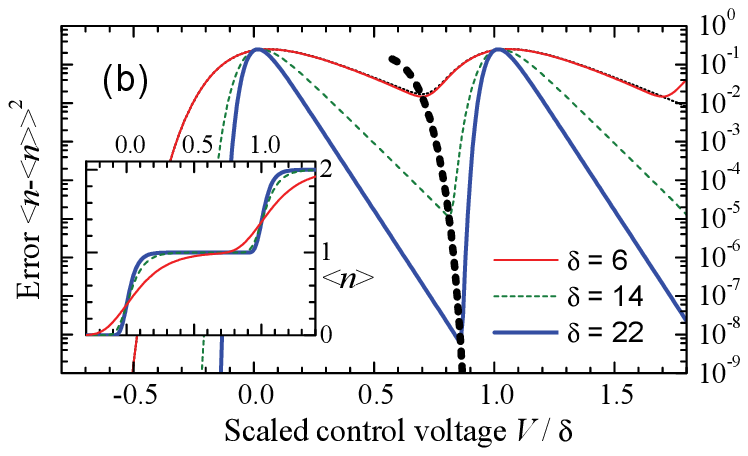}
  \caption{(color online)
   (a) Overview of the model. (b) Main results: 
   the variance, $\qav{n-\qav{n}}^2$, and the average (inset),
   $\langle n \rangle$, of the number of captured electrons $n$,
    for $\alpha_n=1$ ,   $\delta_1=0$ and $\delta_n= \delta$
    as functions of control voltage $V$ (in units of $\delta$).
   All results are calculated using $N=5$ and Eq.~\eqref{eq:generalX}, except
   for the thin dotted line that shows Eq.~\eqref{eq:3eqs} with $N=2$.
   The thick dotted line traces the position and the value
   of variance as $\delta$ is varied.
   }
  \label{fig:newfig1}
\end{figure}

\paragraph*{Decay cascade model.}
The initialization process is shown pictorially
in Fig.~\ref{fig:newfig1}a.
The relaxation rate of electrons in the QD,
$\Gamma_{\text{RC}}=(RC)^{-1}$, is reduced at
a characteristic speed
$\beta = |\dot{\Gamma}_{\text{RC}}/\Gamma_\text{RC}|$.
(Here  $R$ and $C$ are the resistance and the capacitance
of the QD with respect to its environment, respectively.)
Simultaneously, $\varphi(t)$ grows more negative.
Ejection of electrons at rate  $\Gamma_\text{ad}=| e \dot{\varphi}/E_c|$
is required for
the electron number distribution $P_n(t)$ to stay close to  instantaneous equilibrium
($E_c=e^2/C$ is the charging energy and  $e$ is the electron charge).
Such
adiabatic following becomes impossible due to insufficient escape rate
at times
beyond the non-adiabatic crossover
time $t_0$ defined by $\Gamma_{\text{RC}}(t_0) = \Gamma_{\text{ad}}$.
Crucially for our scheme, if $\Gamma_{\text{ad}}$ can be made much larger than $\beta$
then excitation from the Fermi sea \cite{Flensberg1999} can be ignored while
the QD remains loose enough at $t>t_0$ to ``forget''
the adiabatic initial condition $P_n(t_0)$ and yield eventual accuracy
dictated by the decay cascade mechanism.

For $t > t_0$ we write down a general kinetic equation
\begin{align} \label{eq:mastereqsX}
  d P_n(t)/d t & = -\Gamma_n(t) \,  P_n(t) + \Gamma_{n+1}(t) \,  P_{n+1}(t) \, , \\
  P_n(t_0) &= \delta_{n,N}   , \; \lim_{t \to \infty} \Gamma_n(t) =0 \, ,
  \label{eq:ini}
\end{align}
where $\Gamma_n$ is the decay rate of the charge state with $n$ electrons on the QD.
The empty dot ($n=0$) is an absorbing state, $\Gamma_0 \equiv 0$, and
the distribution is normalized, $\sum_{n=0}^{\infty} P_n=1$.
$N$ is
the initial number of electrons and $\Gamma_{N}(t_0)$
is identified with $\Gamma_{\text{ad}}$  (see Fig.~\ref{fig:newfig1}a).
The use of Eqs.~\eqref{eq:mastereqsX}--\eqref{eq:ini}
to describe QD initialization assumes
(i) randomization of the microstate
corresponding to a given $n$ on a time scale $\tau \ll
\Gamma_n^{-1}$ so that Markov approximation is justified and
the transition rates $\Gamma_n$ are well-defined,
(ii) no additional loading of electrons into the QD after $t_0$,
and (iii) sharp initial condition, Eq.~ \eqref{eq:ini}
(the latter is non-essential).
The system of equations \eqref{eq:mastereqsX} is rather general,
and has also been used in the context of dynamic QD evolution
\cite{astley2007,miyamoto2008} (although not for the initialization stage).

A general iterative solution to Eq.~\eqref{eq:mastereqsX} is 
\begin{align} \label{eq:ingralform}
  P_n(t) &  = \int_{t_0}^{t} \! e^{
  - \int_{t'}^t \Gamma_{n}(\tau) \, d \tau } \Gamma_{n+1}(t') \,  P_{n+1}(t')  \, dt' \, ,
\end{align}
where $P_{N+1}(t)  = \delta(t-t_0)/ \Gamma_{N+1}(t_0)$ ensures fulfillment
of the initial condition \eqref{eq:ini}.
Our strategy is, firstly, to identify the general properties of
Eq.~\eqref{eq:ingralform} that result
in a well-defined final electron number, and, secondly,
to introduce more specific
physical assumptions that connect the model with potential
experimental realizations.

\paragraph*{Conditions for accurate initialization.}
Consider an additional assumption
(which can be relaxed later) that
time-dependence of $\Gamma_n(t)$ is the same for all
 $n$, namely, $\Gamma_{n}(t)/\Gamma_{n-1}(t) \equiv
e^{\delta_n}=\text{const}$. In this case the final ($t \to \infty$)
distribution $P_n$ depends only on the dimensionless integrals
  $X_n \equiv \int_{t_0}^{\infty} \Gamma_n(t) \, d t$
and is given by
\begin{align}
    P_n(\infty) & = \sum\nolimits_{k=n}^{N} Q_{nk} \,
   C_k \,  e^{-X_k} \, , \label{eq:generalX} \\
  C_k  & =  -\sum\nolimits_{m=k+1}^{N} C_m Q_{km}  \, ; \; C_N=1 \, , \\
  Q_{nk} & =   \prod\nolimits_{m=n}^{k-1}  \frac{X_{m+1}}{X_{m}-X_k}
     \, ; \; Q_{nn} = 1  \, . \label{eq:Q}
\end{align}
The solution \eqref{eq:generalX} is rather
unilluminating, but can be investigated numerically
(see below). For now let us focus on the limit of
cascade timescale separation:
\begin{align} \label{eq:sepcascades}
  \ldots \gg X_{n+1} \gg  X_n \gg X_{n-1} \gg \ldots \, .
\end{align}
In this limit Eq.~\eqref{eq:Q} simplifies to $Q_{k-1,k}\!=\!-1$ and
$Q_{nk}\!=\!0$ for all $n\!<\!k-1$. This in turn means that  $C_n\!=\!1$
for all $ 0\!\le\!n\!\le\!N$. The solution becomes $P_N(\infty) =
e^{-X_N}$ and
\begin{align} \label{eq:stacking}
  P_n(\infty) = e^{-X_{n}} - e^{-X_{n+1}} \text{ for } 0<n<N \, .
\end{align}
Our model is justified for $\Gamma_N(t_0) \gg |\dot{\Gamma}_N/\Gamma_N|$,
which implies $X_N \gg 1$ (for smoothly decaying rates).
Thus there will be such integer $n_0 < N$ that
\begin{align} \label{eq:justn0}
   X_{n_0+1} > 1 >   X_{n_0}   \text{ and } X_{n_0+1} \gg X_{n_0} .
\end{align}
Under Eq.~\eqref{eq:sepcascades} it is sufficient to consider
only three probabilities
to be non-zero:
\begin{multline}
\{ P_{n_0+1}, P_{n_0},  P_{n_0-1}  \} = \\
\{ e^{-X_{n_0+1}}, e^{-X_{n_0}}-e^{-X_{n_0+1}},
1-e^{-X_{n_0}} \} \, .\label{eq:3eqs}
\end{multline}
We see that the probability distribution is dominated by $P_{n_0} \to 1$
if $X_{n_0+1} \gg 1 $ and $X_{n_0} \ll 1 $.
The meaning of this condition
is simple: the state with $n_0+1$ electrons
is unstable enough to have decayed 
 while $n_0$ is stable 

Time-scale separation expressed by Eq.~\eqref{eq:sepcascades}
can be taken into account directly in the most
general solution \eqref{eq:ingralform}, without requiring same
time-dependence of $\Gamma_n(t)$'s.
The mathematics of this derivation can be summarized as follows:
(a) observe that $\Gamma_{n+1} P_{n+1}\!=\!-\frac{d}{dt} \sum_{m>n}\!P_{m}$ exactly;
(b) assume that all $P_m(t)$'s with $m>n$ compared to $P_n(t)$ itself vary on a time-scale much closer to $t_0$,
so that $\Gamma_{n+1}(t') P_{n+1}(t') \propto \delta(t_0\!-\!t')$ in Eq.~\eqref{eq:ingralform};
(c) solve the resulting $ P_{n}(\infty)\!=\!e^{-X_n}(1\!-\!\sum\nolimits_{m=n+1}^N P_m) $
to get
\begin{align} \label{eq:branchingsolution}
  P_{n} (\infty) 
  & = e^{-X_n} \prod\nolimits_{m=n+1}^{N} (1-e^{-X_m}) \, .
\end{align}
Finally, examining Eq.\eqref{eq:branchingsolution}
under condition \eqref{eq:justn0} reveals difference
from Eq.~\eqref{eq:3eqs} of at most
$X_{n_0}$ which is reached at $ X_{n_0} \ll X_{n_0+1} \approx 1$.

\paragraph*{Parametric control of decay rate hierarchy.}
 Usefulness of Eq.~\eqref{eq:generalX} is limited unless
 dependence of $X_n$'s on commonly accessible
 external control parameters
 can be established. Typically, escape rates
 depend exponentially on the height of the confining barrier,
 which in turn can be controlled by a gate voltage $V$.
 Tuning the decay rates $\Gamma_n(t)$ by $V$ would affect
 all decay rates simultaneously. Thus we propose
 \begin{align} \label{eq:exponential}
   \ln X_n = -\alpha_n V + \sum\nolimits_{i=1}^{n} \delta_i \, .
\end{align}
Here $\alpha_n$ and $\delta_{n}$ are phenomenological constants
that can be readily extracted experimentally as discussed below.

Figure~\ref{fig:newfig1}b
shows the behavior of the first two moments
for equal and fixed $\alpha_n=1$ and $\delta_n = \delta$
($\delta_1=0$ is a mere shift of $V$). For $\delta>6$
the difference between the approximations \eqref{eq:generalX},
\eqref{eq:3eqs} and \eqref{eq:branchingsolution} is negligible
and Eq.~\eqref{eq:3eqs} is sufficient to describe the $n_0$-th
step of a staircase $\qav{n}(V)$ regardless of $N$.
The minimal variance $\qav{n-\qav{n}}^2$ is achieved
at optimal values of $V$ that are easily found from Eqs.~\eqref{eq:3eqs} and
\eqref{eq:exponential}. The minimal value of $1-P_1(V)$
decays roughly exponentially with $\delta$.

\paragraph*{Possible experimental realizations.}
We expect to find good candidates for experimental
realization of the proposed initialization mechanism
among dynamic-QD-based electron pumps since
backtunneling relaxation has been
found~\cite{Aizin1998,Leicht2009} to
take place during certain parts of the pump cycle.
The pumped current $I_\text{pump}$ consists of electrons
captured from the source
and subsequently ejected into the drain.
It can be related to model prediction via
\begin{align}\label{eq:current}
  I_\text{pump} = e f \qav{n} \equiv e f \sum\nolimits_n n P_n(\infty)
\end{align}
($f$ is the frequency
which the pump cycle is repeated) provided that
(a) ejection to the drain
starts after the escape back to the source has stopped,
and (b) the ejection is complete.
In some experimental settings there is strong
evidence that the latter condition can be
ensured~\cite{kaestner2008,miyamoto2008} while in others
\cite{Aizin1998,Robinson2001} conditions (a) and (b) have been
conjectured based on electrostatic modeling.

\begin{figure}
 \includegraphics[width=7cm,clip]{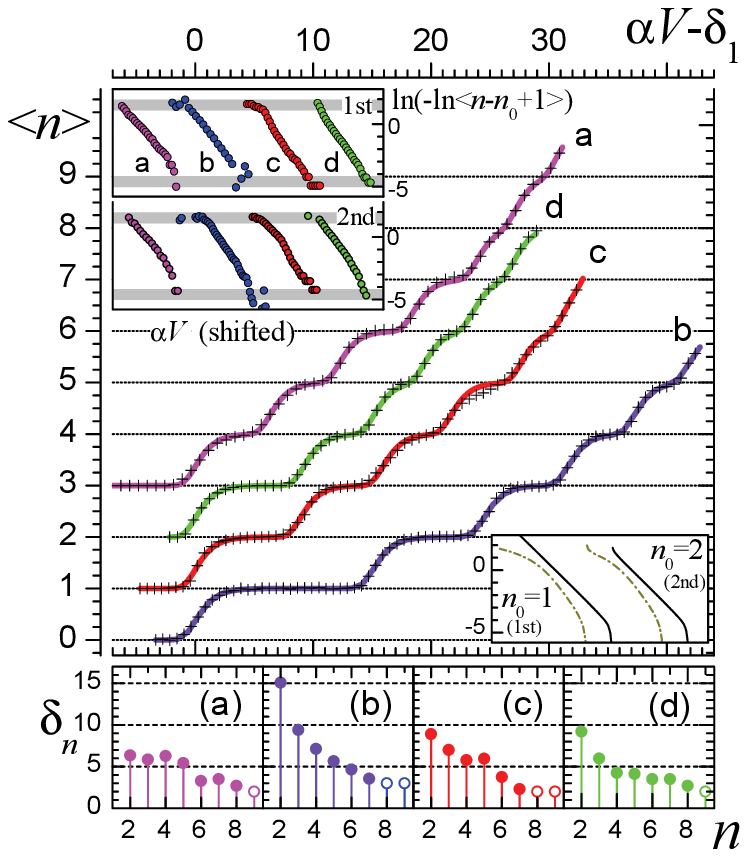}
  \caption{(color online)
Upper panel, main graph: decay cascade model fits (solid line)
compared to measured current $I/(ef)$ (pluses)
from Fig.~2a of Ref.~\onlinecite{fujiwara2008} (a, magenta), Fig.~1a of Ref.~\onlinecite{kaestner2009a} (b, blue), and Fig.~1 of
Ref.~\onlinecite{Janssen2001} (c, red), and to classical simulation results from
Fig.~6 of Ref.~\onlinecite{Robinson2001} (d, green).
The graphs are shifted vertically for clarity. Insets are discussed in the text.
Lower panels:
fitted values of $\delta_n$ for each case.
}
  \label{fig:Figure_comparison}
\end{figure}

We have used the ansatz \eqref{eq:exponential}
and the solution \eqref{eq:generalX} in Eq.~\eqref{eq:current} to fit
experimental data from various electron pump
devices~\cite{fujiwara2008,kaestner2009a,Janssen2001},
the results are shown in Fig.~\ref{fig:Figure_comparison}.
In all cases $\alpha_n$ were found to vary weakly with $n$,
so we have set $\alpha_n=\alpha$ constant for the entire voltage range for each device.
Data set ``a'' in
Fig.~\ref{fig:Figure_comparison} was obtained in
Ref.~\onlinecite{fujiwara2008} for a silicon nanowire
metal-oxide-semiconductor field-effect transistor (FET) driven
by voltage pulses, while the data labelled ``b'' correspond to
a sinusoidal modulation of an AlGaAs/GaAs nanowire
metal-semiconductor FET in the quantum Hall
regime~\cite{kaestner2009a}.
Yet another realization is exemplified by the data
obtained in Ref.~\onlinecite{Janssen2001},
for an AlGaAs/GaAs split-gate device (``c'' in
Fig.~\ref{fig:Figure_comparison}), where the time
dependent potential was generated by SAWs.
The voltage applied to the
split gate superimposes the SAW-potential and
tunes the height of the confining barriers corresponding to the
control voltage $V$.

Figure~\ref{fig:Figure_comparison} demonstrates particular advantage
of $\alpha_n\!=\!\alpha$: $(n\!-\!1)$-th step in $I_{\text{pump}}(V)$ has the length
of $\delta_n$ on the scale of $\alpha V\!-\!\delta_1$. Thus the set of voltage-independent
 $\delta_n=\ln X_n/X_{n-1}$ can be used a fingerprint of a particular device,
 see the lower panels in  Fig.~\ref{fig:Figure_comparison}.
Plotting $\ln (-\ln \qav{n-n_0+1})$ as a function of $V$ serves
 as a quick test of the ansatz \eqref{eq:exponential}
 since the plateaux part dominated by $X_{n_0}$ must
 show up as a straight line.
 This is demonstrated in the lower right inset of
 Fig.~\ref{fig:Figure_comparison}  for the analytic $\qav{n}$ with $\delta=6$
 (same parameters as in Fig.~\ref{fig:newfig1}b) and in the upper left inset
 for the empirical data (same data sets as in the main panel).
 For contrast, we show by a green dash-dotted line an \emph{ad hoc} fit with a sum
 of symmetric Fermi-like step functions; the deviation from the cascade model is notable.

\begin{figure}
 \includegraphics[width=5.8cm,clip]{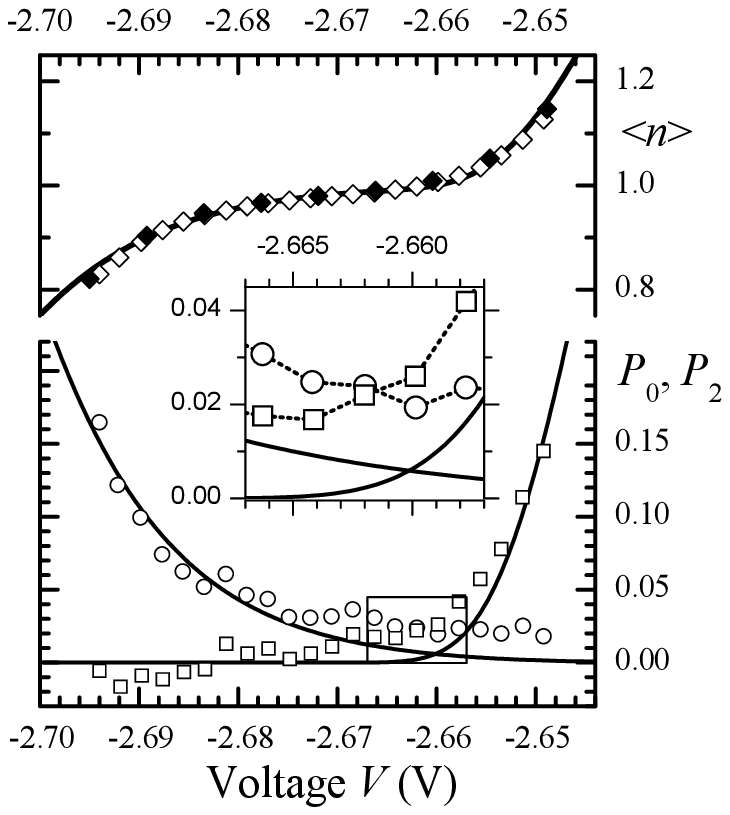}
    \caption{Comparison to shot noise measurements on a SAW pump taken from Fig.~4a of Ref.~\onlinecite{Robinson2005}.
    Upper panel:  measured $I_{\text{pump}}/(ef)$ [$\blacklozenge$ from $I_{\text{pump}}(V)$,
    $\lozenge$ from $P_0(V)$ and $P_2(V)$] and calculated $\qav{n}$ (solid line).
    Parameters fitted in  Eq.~\eqref{eq:generalX} with $N=2$ and $\alpha_1=\alpha_2$:
    $\alpha_1=92.04 \, \text{V}^{-1}$, $\delta_1=-240.7$ and $\delta_2=6.252$.
    Lower panel and central inset: experimental probabilities
    $P_0$ ($\bigcirc$) and $P_2$ ($\square$) compared to Eq.~\eqref{eq:generalX}
    (solid line) with parameters derived from the fit to $I_{\text{pump}}$.
\label{fig:noise}}
\end{figure}

Our results for the second moment can be tested
by measuring  the low-frequency shot noise power
spectral density \cite{Galperin2001}
\begin{equation} \label{eq:noisedef}
  S = 2 \, e^2 f \, \bigl (\langle {n^2} \rangle -\qav{n}^2 \bigr ) \, .
\end{equation}
Beyond the conditions (a) and (b) discussed above, Eq.~\eqref{eq:noisedef}
assumes \cite{Galperin2001} that (c) the temporal width $ d\tau $  of the electron ejection current
pulse  is much less than the repetition period, $d\tau
\, f \! \ll \! 1$. This regime has been probed experimentally by
Robinson and Talyanskii~\cite{Robinson2005}, 
and we find good agreement with their data, see Fig.~\ref{fig:noise}.

\paragraph*{Discussion and outlook.}
From the observations demonstrated in
Figs.~\ref{fig:Figure_comparison} and \ref{fig:noise} we
conclude that the decay cascade initialization scheme
can be readily implemented experimentally.
Several ways of controlling $\delta_n$, thus achieving more accurate
initialization can be
suggested: (a) tighter confinement to increase the charging
energy, (b) employing more energy-selective
barriers~\cite{Kaestner2007c}, or (c) lowering
the local temperature. Suggestion (c) applies when escape
is determined by thermal activation.
Note that this classical regime was investigated in electron dynamics
simulations of Robinson and Barnes \cite{Robinson2001}, see trace
``d'' in Fig.~\ref{fig:Figure_comparison}, which demonstrates
the universality of the decay-cascade model.
Within a classical independent electron picture,
$\delta_n$ would be controlled by the difference in barrier height
for the most energetic electron, $\delta_n \approx a
(E_{n}-E_{n-1})$ where  $E_n$ is the ionization energy of a QD
with $n$ electrons, and $a$ is an inverse effective temperature
in this picture. The experiment of Ref.~\onlinecite{fujiwara2008}
may have operated under these conditions since the reported
temperature $T=20\,$K is relatively high. The value of
$\delta_2=6.4$ from Fig.~\ref{fig:Figure_comparison}
then gives $E_C=11\,$meV in agreement the experimental estimate \cite{fujiwara2008}
$E_C=10\,$meV.
In other implementations
lowering the temperature of the device may be not that effective:
comparing SAW-~\cite{Janssen2001} and FET-based
pumps~\cite{kaestner2009a} (traces ``c'' and ``b'' in
Fig.~\ref{fig:Figure_comparison} respectively),
we see a large difference in $\delta_2$ of the first plateaux.
Despite a similarity in QD area, temperature and material implementation
the maximally achievable accuracy according to our model is $\qav{n - \qav{n}}^2 \approx 10^{-3}$
for this particular SAW device versus $10^{-5}$ in the case of the FET.
Assuming that tunneling dominates in the latter case,
$\delta_n$ is expected to scale proportional to the difference
in localization lengths between the ground states with $n$ and
$n-1$ charges.

In conclusion, we have proposed a scheme for accurate initialization
of QDs and analyzed it quantitatively. We have shown that the model
may be readily implemented using non-adiabatic pump architectures.
It allows to extract a device-specific fingerprint which can be used to
predict the results of a high precision measurement from a low precision
characterization. In this way one can efficiently evaluate and
optimize different pump architectures as required for fundamental
metrology and adapt them for dynamic-QD-based quantum information devices.

We thank Ph.\ Mirovsky for
discussions and T.J.B.M.\ Janssen for providing raw data for
Fig.~\ref{fig:Figure_comparison}c. V.K. has been supported by
ESF  project No.2009/0216/1DP/1.1.1.2.0/09/APIA/VIAA/044.
B.K. has been supported by
EURAMET joint research project with European Community's
$7^{\text{th}}$ Framework Programme, ERANET Plus under Grant
Agreement No. 217257.


\end{document}